\documentclass[referee]{raa}           
\usepackage{graphicx,times}
\usepackage{natbib}
\usepackage{amssymb,amsmath}
\usepackage{gensymb}
\bibpunct{(}{)}{;}{a}{}{,}

\usepackage{xcolor}
\usepackage{ulem}

\usepackage{array}
\newcommand{\PreserveBackslash}[1]{\let\temp=\\#1\let\\=\temp}
\newcolumntype{C}[1]{>{\PreserveBackslash\centering}p{#1}}
\newcolumntype{R}[1]{>{\PreserveBackslash\raggedleft}p{#1}}
\newcolumntype{L}[1]{>{\PreserveBackslash\raggedright}p{#1}}

\usepackage[a4paper=true,pagebackref=true]{hyperref}
\hypersetup{pdftitle = A photometric study of two neglected  eclipsing binaries, 
pdfauthor = Kudak V. Fedurco M. Perig V. Parimucha \v{S}., 
pdfsubject= EB stars, 
pdfkeywords = binaries: close --- binaries } 
\hypersetup{colorlinks = true, linkcolor = green, anchorcolor = red, citecolor = blue, filecolor = red, pagecolor = red, urlcolor = red}

\begin{document}

   \title{A photometric study of two neglected  eclipsing binaries}
   
 \volnopage{ {\bf 20XX} Vol.\ {\bf X} No. {\bf XX}, 000--000}
   \setcounter{page}{1}

   \author{V. Kudak\inst{1}, {M}. Fedurco\inst{2}, V. Perig\inst{1}, \v{S}. Parimucha\inst{2}}
   
\institute{Laboratory of space researches, Uzhhorod National University, Uzhhorod, Daleka Str., 2A,  88000, Ukraine; {\it e-mail:~lab-space@uzhnu.edu.ua}\\
        \and
          Institute of Physics, Faculty of Science, University of P.J. \v{S}af\'arik, Ko\v{s}ice, Park Angelinum 9, 04001, Slovakia; {\it e-mail:~stefan.parimucha@upjs.sk}\\
       \vs\no
   {\small Received~~20xx month day; accepted~~20xx~~month day}}

\abstract{ 
We present the first BVR photometry, period variation, and photometric light-curve analysis of two poorly studied eclipsing binaries V1321 Cyg and CR Tau. Observations were carried out from November 2017 to January 2020 at the observatory of Uzhhorod National University. Period variations were studied using all available early published as well as our minima times. We have used newly developed ELISa code for the light curve analysis and determination of photometric parameters of both systems. We found that V1321~Cyg is a close detached eclipsing system with a low photometric mass ratio of $q=0.28$ which suggests that the binary is a post mass transfer system. No significant period changes in this system are detected. CR~Tau is, on the other hand, a semi-detached system where the secondary component almost fills its Roche lobe. We detected a long-term period increase at a rate of $1.49 \times 10^{-7} d/y$, which support mass transfer from lower mass secondary component to the more massive primary.
}

   \authorrunning{V.I. Kudak et al. }            
   \titlerunning{A photometric study of two neglected  eclipsing binaries}  
   \maketitle

%
\section{Introduction}           
\label{sect:intro}
Eclipsing binaries are an important group of variable stars, where both components are obscured for the observer during their mutual motion around a common centre of mass. They exhibit features in their light curves, which are specific and well recognized among all variable stars. The shape of their light curves depends on the physical properties of the components and geometrical configuration \citep{2001icbs.book.....H, Prsa2019}. Analysis of light-curves of eclipsing binaries can reveal, among others, relative dimensions of stars, their effective temperatures, orbital inclination, the eccentricity of the orbit, and potential spots. Together with radial velocities obtained from spectroscopic observations, we can determine masses of the components, their distances, and radii. 

The shapes of components in binary stars are described by Roche geometry \citep{Prsa2019}. According to this, three configurations of binary systems are possible, detached (both components are in their Roche lobes), semi-detached (one component fill its Roche lobe), and contact, where both components overfill their Roche lobes. All this is reflected in the light curves and have also other observational consequences like a period change due to mass transfer, angular momentum loss \citep[e.g.,][]{Yang2009} and/or magnetic braking \citep{1992ApJ...385..621A}.

In this paper, we present photometry, period, and light-curves analysis of two neglected detached binaries, which were not up to now, studied in more details in literature:

\textbf{V1321 Cyg} (NSVS 5731097) was for the first time mentioned as eclipsing variable in \cite{Romano1967}. In the database of \cite{Kreiner2004}, the period of the system is listed as P\,=\,0.3640924 \,days. \cite{Otero2006} redefined system as Algol-type binary with the orbital period P\,=\,0.72818 \,days. In the catalogue of \cite{Avvakumova2013}, the orbital period of the system was again set to half of the previous value. The distance to the system is 735$\pm$12 pc according to GAIA DR2 release \citep{2018AJ....156...58B}.

\textbf{CR Tau} (GSC 01862-01633) was discovered by \cite{Hoffmeister1949} who also determined ephemeris from minima times from photographic plates. The system was neglected till the paper from \cite{Agerer1999}, who presented the first CCD light-curve of the system and determined new ephemeris with period P\,=\,0.6827035\,days. It is included in catalogue of Algol-type eclipsing binaries from \cite{2004A&A...417..263B} and in the catalogs \cite{Malkov2006} and \cite{Avvakumova2013}. This eclipsing binary was also monitored by the OMC instrument (The Optical Monitoring Camera) on-board INTEGRAL satellite which provided photometry measurements in the Johnson V-band \citep{Alfonso2012}. The distance to the system was established to 796$\pm$26 pc according to GAIA DR2 release \citep{2018AJ....156...58B}.

\section{Observations and data reduction}
Observations of all studied eclipsing binary systems were carried out at Derenivka Observatory of Uzhhorod National University, Ukraine (Lat: 48.563417 N; Long: 22.453758 E). Measurements were collected from November 2017 to January 2020. 
For our observation, we have used a Newton-type telescope with a diameter of 400 mm and a focus of 1750 mm. It is accompanied by FLI PL9000 CCD camera (array 3056x3056, pixel size 12$\mu m$) with Johnson $BVR$ photometric filters. The field of view of such configuration of the system is 1.21\degree x 1.21\degree. The journal of our CCD observation is given in Table~\ref{tab:obs-log}.

\begin{table}[t]
\bc
\begin{minipage}[]{0.8\linewidth}
\centering
\caption[]{The journal of CCD photometric observations. Phase is calculated according to ephemeris determined in Section~\ref{sec:period}. \label{tab:obs-log}}\end{minipage}
\small
\begin{tabular}{ccccc}
\hline \hline
\textbf{System}     &\textbf{Date} & \textbf{Time(UT)}  &\textbf{Phase}  & \textbf{Filters} \\
\hline
V1321 Cyg  & Nov 05 18 & 16:27 - 23:22  & 0.047 - 0.442 & B, V, R   \\
           & Nov 06 18 & 16:32 - 19:55  & 0.426 - 6.618 & B, V, R   \\
           & Nov 15 18 & 19:59 - 21:31  & 0.977 - 0.053 & B, V, R   \\
           & Oct 15 19 & 17:09 - 23:56  & 0.500 - 0.882 & B, V, R   \\
           & Oct 17 19 & 16:39 - 00:19  & 0.213 - 0.635 & B, V, R   \\
           & Oct 22 19 & 16:54 - 01:02  & 0.093 - 0.544 & B, V, R   \\
           & Oct 27 19 & 16:57 - 18:47  & 0.961 - 0.068 & B, V, R   \\
           & Jan 15 20 & 16:40 - 20:04  & 0.808 - 0.971 & B, V, R   \\
\hline
CR Tau     & Nov 30 18 & 18:31 - 04:31  & 0.945 - 0.562 & B, V, R \\ 
           & Oct 25 19 & 23:13 - 03:24  & 0.146 - 0.401 & B, V, R \\ 
           & Oct 27 19 & 20:45 - 04-02  & 0.927 - 0.371 & B, V, R \\ 
           & Nov 30 19 & 20:05 - 03:07  & 0.688 - 0.117 & B, V, R \\ 
           & Jan 16 20 & 19:37 - 21:43  & 0.503 - 0.634 & B, V, R \\ 
\hline\hline
\end{tabular}
\ec
\end{table}

The CCD images were reduced in the usual way (bias and dark subtraction, flat-field correction) using software package CoLiTecVS \citep{Savanevych2017, Parimucha2019}. This package was also used for aperture photometry, calculation of differential magnitudes according to artificial comparison star as well as calibration to the standard photometric
system. The comparison stars used for the determination of artificial ones were selected manually according to the similarity of the studied binaries (brightness,  distance on the sky). This approach significantly improves the quality of photometric measurements. Due to not a stable night-to-night observing conditions, the average precision of our measurements reached $\sim$0.02 mag in V and R filters and $\sim$0.04 mag in B filter, respectively for CR Tau. Similarly, for the fainter binary V1321 Cyg, the average precision of individual measurements is a little worse, $\sim$0.03 mag in V and R filters and $\sim$0.05 mag in B filter, respectively. The comparison stars used in our study together with their magnitudes from the NOMAD catalogue \citep{Zacharias2004, Zacharias2005} are listed in Table~\ref{tab:comp_stars}.

\begin{figure}[t]
	\centering
	\includegraphics[width=0.80\textwidth, angle=0]{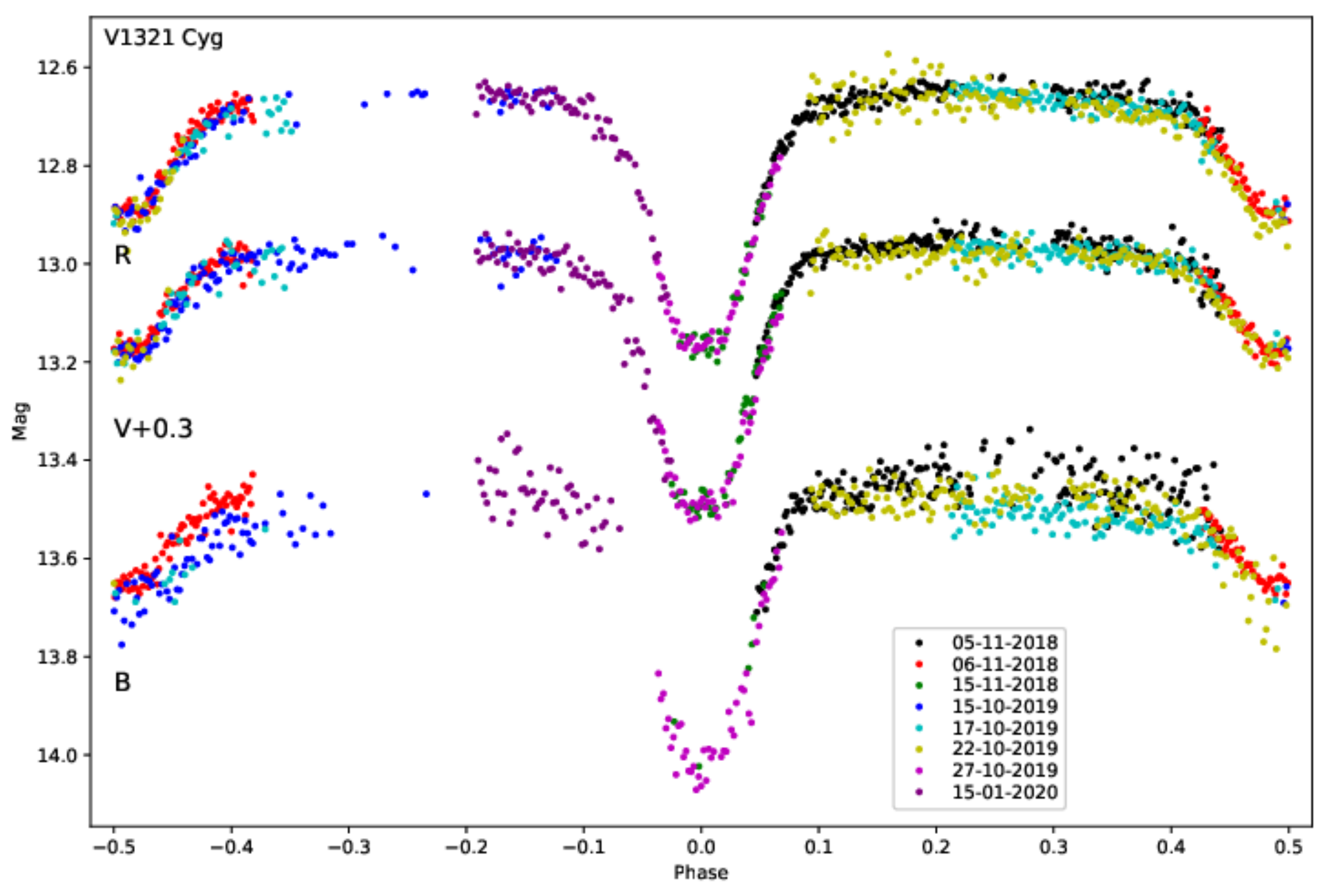}
	\includegraphics[width=0.80\textwidth, angle=0]{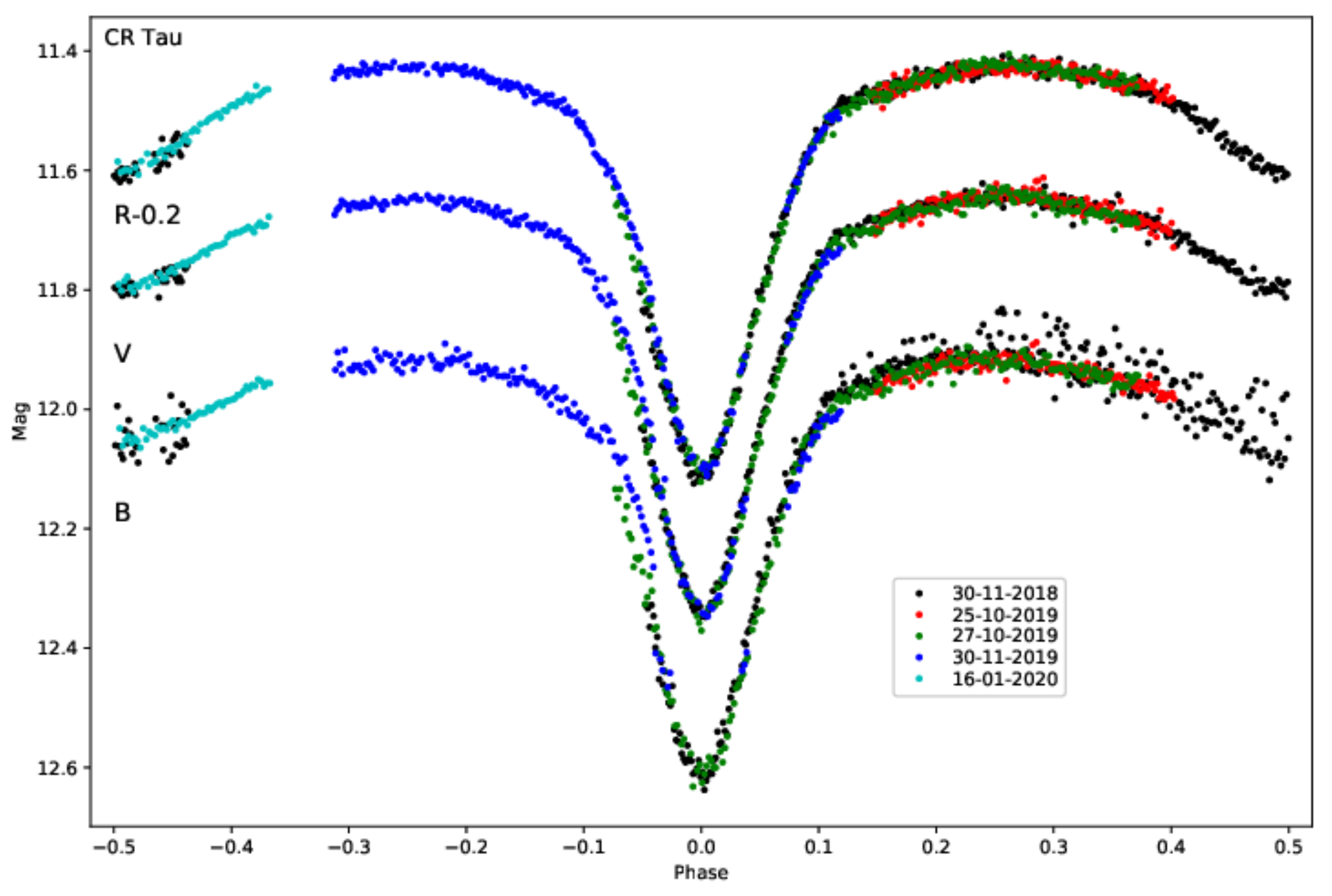}

	\caption{Phased light curves of V1321 Cyg (top) and CR Tau (bottom) systems in B,V,R passbands by dates of observations.}
	\label{fig:phot_all}
\end{figure}

The resulting light-curves of all eclipsing binaries are depicted in Fig.~\ref{fig:phot_all}. The light-curves were phased according to ephemeris determined from O-C variations analyzed in the next chapter.

\begin{table}[b]
\bc
\begin{minipage}[]{0.8\linewidth}
\centering
\caption[]{Comparison stars used for a determination of artificial comparison stars. BVR magnitudes are taken from the NOMAD Catalogue \citep{Zacharias2004, Zacharias2005} \label{tab:comp_stars}}\end{minipage}
\setlength{\tabcolsep}{1pt}
\small
\begin{tabular}{L{1.8cm}|C{2.7cm}|C{1.8cm}C{2cm}|C{1cm}C{1cm}C{1cm}}
\hline \hline
\textbf{System}     &   \textbf{Comparison stars}& \multicolumn{2}{c|}{\textbf{Coordinates}}&\textbf{B}&\textbf{V}&\textbf{R}\\
\textbf{} &   \textbf{NOMAD}          &  $\alpha(2000)$ & $\delta(2000)$ &  &  & \\
\hline\hline
V1321 Cyg  & 1315-0399331 & 20:23:35.66 & +41:32:52.8 & 14.780 & 14.330 & 15.240\\ 
           & 1315-0399475 & 20:23:48.84 & +41:31:36.3 & 12.590 & 12.380 & 11.460\\
           & 1314-0397913 & 20:23:33.94 & +41:27:56.5 & 14.190 & 13.680 & 14.240\\
           & 1315-0399386 & 20:23:41.82 & +41:35:23.5 & 13.160 & 12.810 & 13.380\\
        `  & 1315-0399546 & 20:23:55.19 & +41:30:28.6 & 15.140 & 14.520 & 14.950\\
\hline
CR Tau & 1140-0090488 & 05:51:17.48&+24:04:56.1& 12.456& 12.072& 10.95\\
       & 1140-0090531 & 05:51:21.78&+24:04:31.5& 12.500& 12.178& 11.91\\
       & 1140-0090614 & 05:51:32.65&+24:04:55.0& 14.813& 14.101& 13.41\\
       & 1141-0092118 & 05:51:38.19&+24:06:58.2& 14.642& 13.073& 11.77\\
\hline \hline
\end{tabular}
\ec
\end{table}
  
\section{Period changes analysis }
\label{sec:period}
The study of period changes of both systems was carried out using their O-C diagrams. In our analysis we have used all available published minima times as can be found in the O-C gateway\footnote{\texttt{http://var2.astro.cz/ocgate/}}, minima times determined from our observations (weighted averages from B,V,R light curves), minima times determined from available SuperWASP data \citep{Pollacco2006} and INTEGRAL-OMC observations \citep{Alfonso2012}. Our new minima times were calculated using the phenomenological method described in \cite{Mikulasek2015}. This method gives a realistic and statistically significant error in determining minima times. Newly calculated minima times are listed in Tab.~\ref{tab:minima}.

\begin{table}
  \bc
  \caption{New times of minima of studied objects. BVR - weighted average from our B,V,R light curves, SWASP - SuperWasp minima, OMC - INTEGRAL-OMC minima. The errors of minima times are given in parenthesis.}
  \label{tab:minima}
  \begin{center}
  \setlength{\tabcolsep}{5pt}
    \begin{tabular}{lc||lc||lc}
    \hline\hline
    HJD(2400000+) & Filter &  HJD(2400000+) & Filter &  HJD(2400000+) & Filter   \\ 
    \hline
    \multicolumn{2}{c||}{\textbf{V1321 Cyg}} & 54371.4324(10)   & SWASP            & 54070.6860(10)& SWASP  \\                
    54278.5914(17)   & SWASP                 & 54394.3668(18)   & SWASP            & 54118.4750(10)& SWASP  \\           
    54279.6808(9)    & SWASP                 & 54398.3755(12)   & SWASP            & 54179.5762(8) & OMC    \\             
    54282.5926(9)    & SWASP                 & 58429.2438(4)    &  BVR             & 54141.3452(3) & SWASP  \\  
    54298.6130(6)    & SWASP                 & 58774.4070(9)    &  BVR             & 54142.3747(9) & SWASP  \\      
    54318.6419(10)   & SWASP                 & 58779.4987(4)    &  BVR             & 54143.3917(5) & SWASP  \\     
    54335.3949(22)   & SWASP                 & 58784.2343(4)    &  BVR             & 54145.4408(9) & SWASP  \\                          
    54337.5702(21)   & SWASP                 & \multicolumn{2}{c||}{\textbf{CR Tau}} & 55070.5057(2) & OMC    \\                            
    54339.3939(11)   & SWASP                 & 54030.7452(5) & SWASP               & 55454.1833(6) & OMC    \\                            
    54340.4820(29)   & SWASP                 & 54050.5426(9) & SWASP               & 55455.8888(2) & OMC    \\                            
    54344.4887(7)    & SWASP                 & 54056.6889(2) & SWASP               & 58453.3040(1) & BVR    \\                            
    54345.5821(29)   & SWASP                 & 54067.6108(4) & SWASP               & 58784.4165(1) & BVR     \\                           
    54363.4224(5)    & SWASP                 & 54069.6594(3) & SWASP               & 58818.5517(3) & BVR     \\      
      \hline\hline
      \end{tabular}
      \ec
  \end{center}
\end{table}

\begin{figure}[tb]
	\centering
	\includegraphics[trim=25 10 35 35, clip, width=0.46\textwidth]{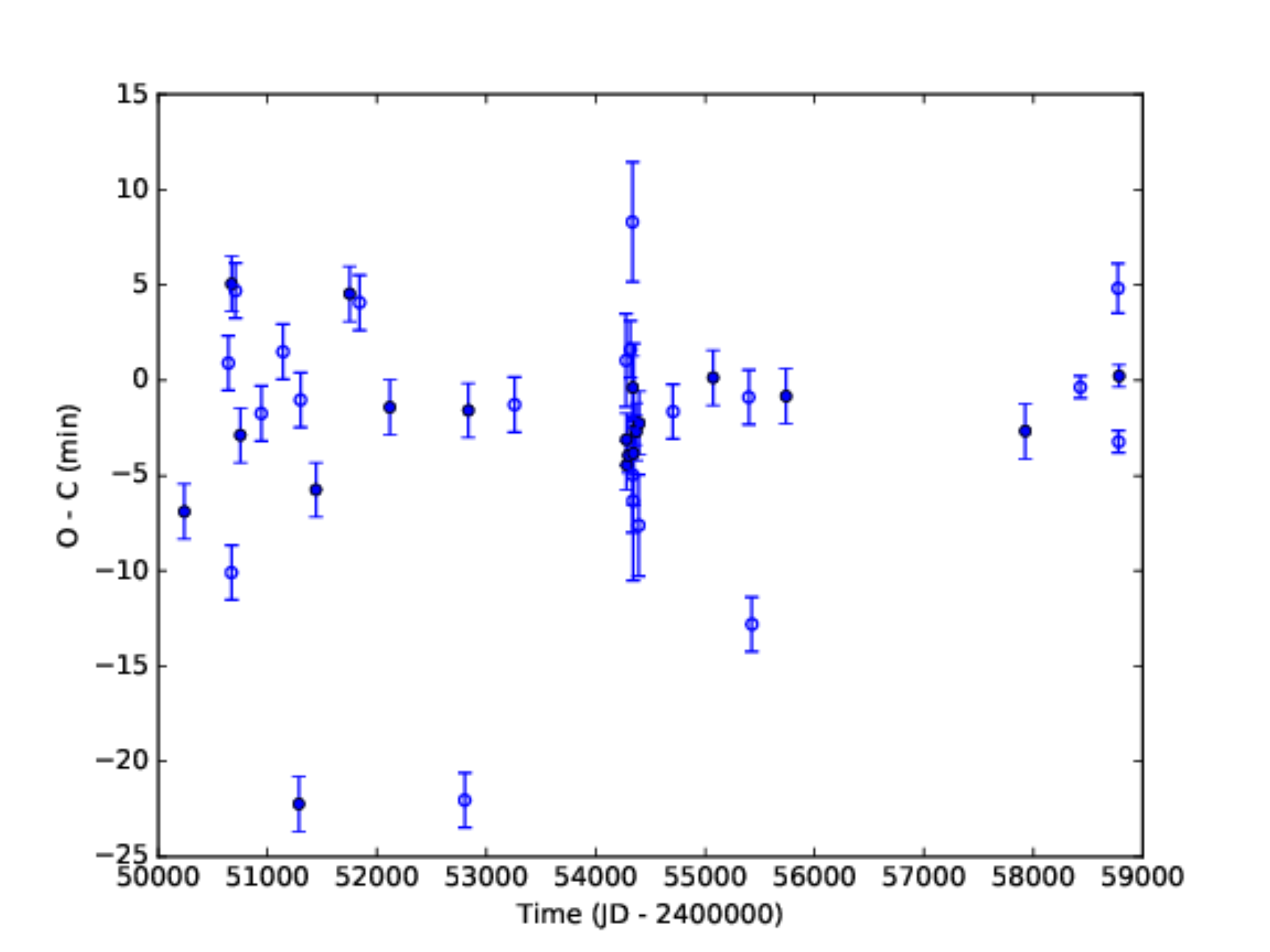}
	\includegraphics[trim=30 10 35 35, clip, width=0.46\textwidth]{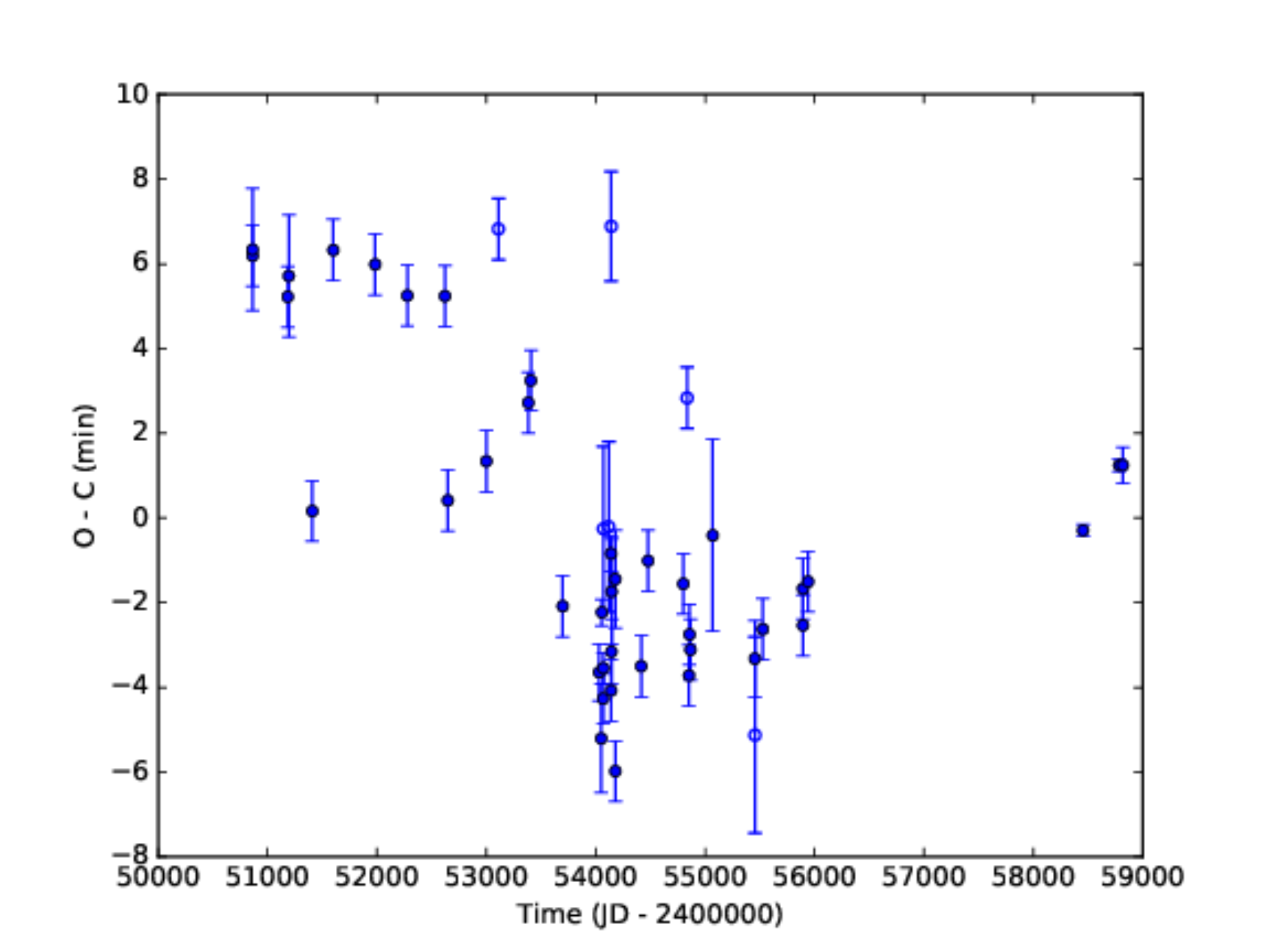}
	\caption{O-C diagrams of V1321 Cyg (left) and CR Tau (right) systems determined from linear ephemeris (\ref{eq:lin_v1321}) and (\ref{eq:lin_cr}). Primary minima are denoted by filled circles and secondary ones by blank circles.}
	\label{fig:oc_both}
\end{figure}

O-C diagram of V1321 Cyg compiled from archived CCD and newly determined ones contains a totally of 88 times of minima. We have omitted old photographic minima times obtained before 1968 because of their large scatter. We also excluded visual observations. The precision of CCD minima times is in the range of $10^{-4}$ days. A weighted least-squares solution using all minima (weights were calculated as $1/\sigma^2$, where $\sigma$ is an error of the minimum) leads to the following linear ephemeris of the system (errors of parameters are given in parenthesis):
\begin{equation}
  \mathrm {Min~I} = \mathrm {HJD~} 2458428.879(3) + 0^{d}.7281849(5) \times E.  
  \label{eq:lin_v1321}
\end{equation}
This ephemeris was used to create the O-C diagram displayed in Fig.~\ref{fig:oc_both}~(left). Despite the fact that we used only CCD observations, quite a large scatter in the resulting diagram is apparent in the range of about 10 minutes. But no significant period changes in this system are detected.

\begin{figure}[tb]
	\centering
	\includegraphics[trim=30 10 35 35, clip, width=0.46\textwidth]{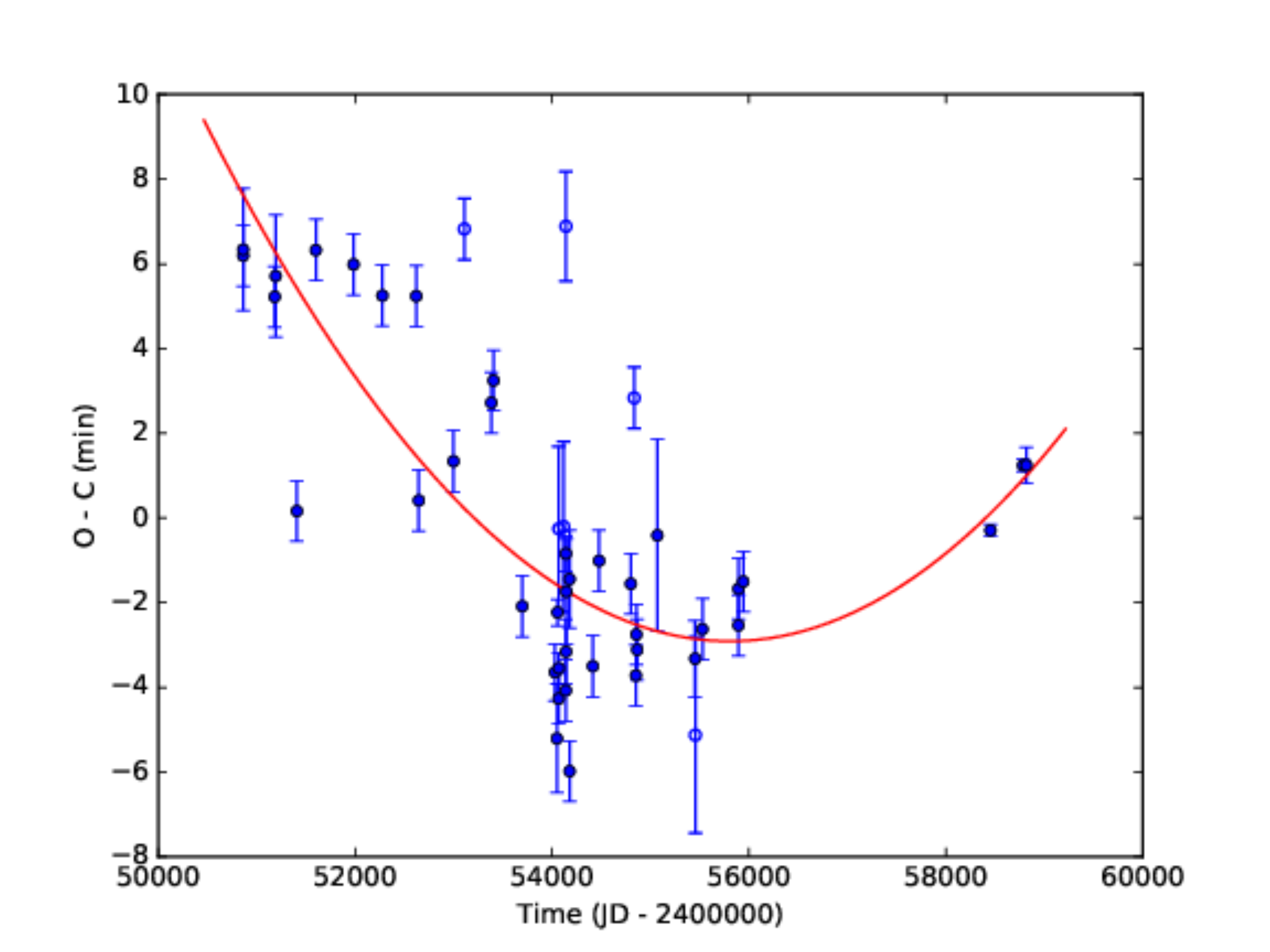}
	\includegraphics[trim=30 10 35 35, clip, width=0.46\textwidth]{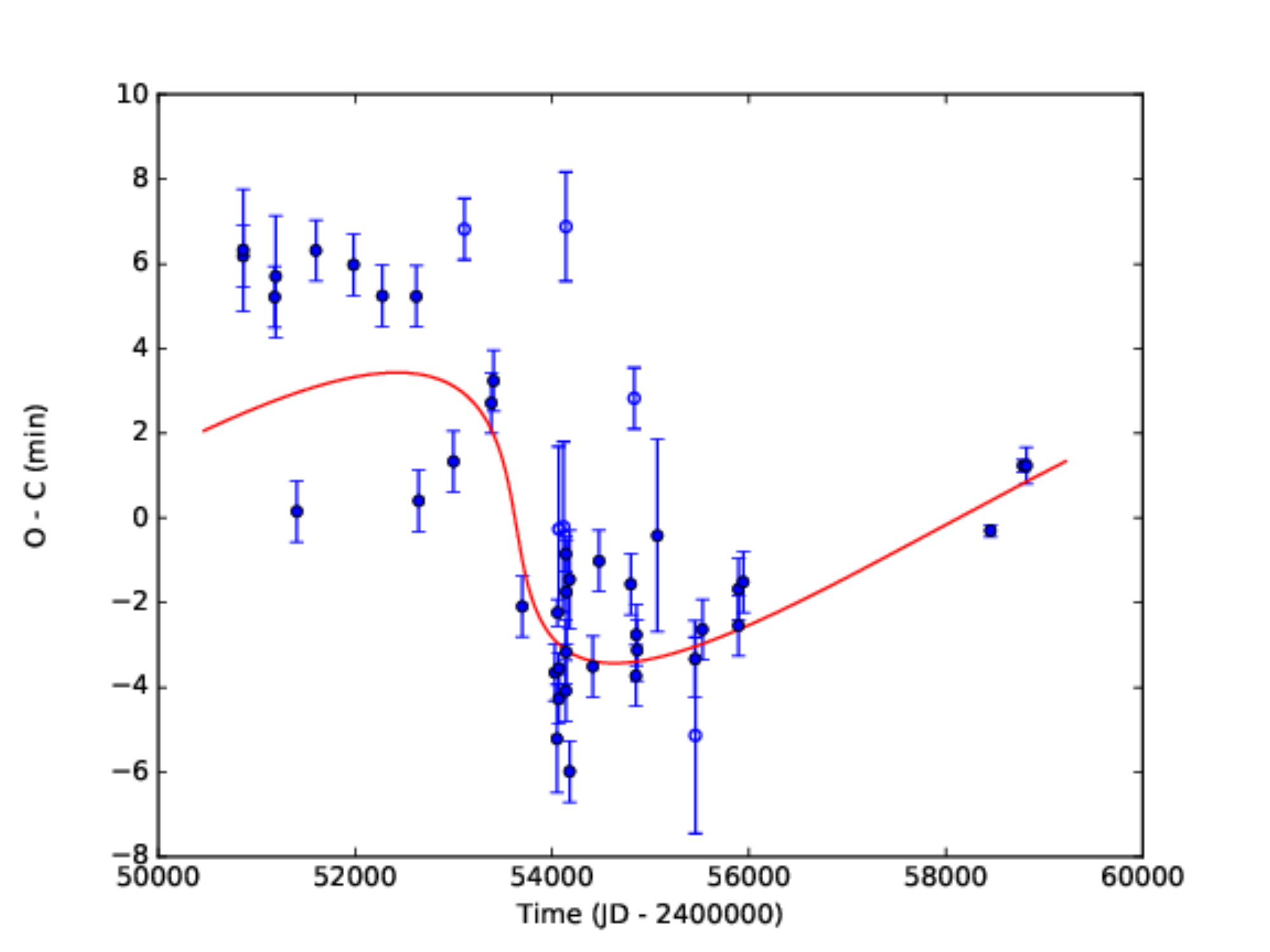}
	
	\includegraphics[trim=30 0 35 15, clip, width=0.46\textwidth]{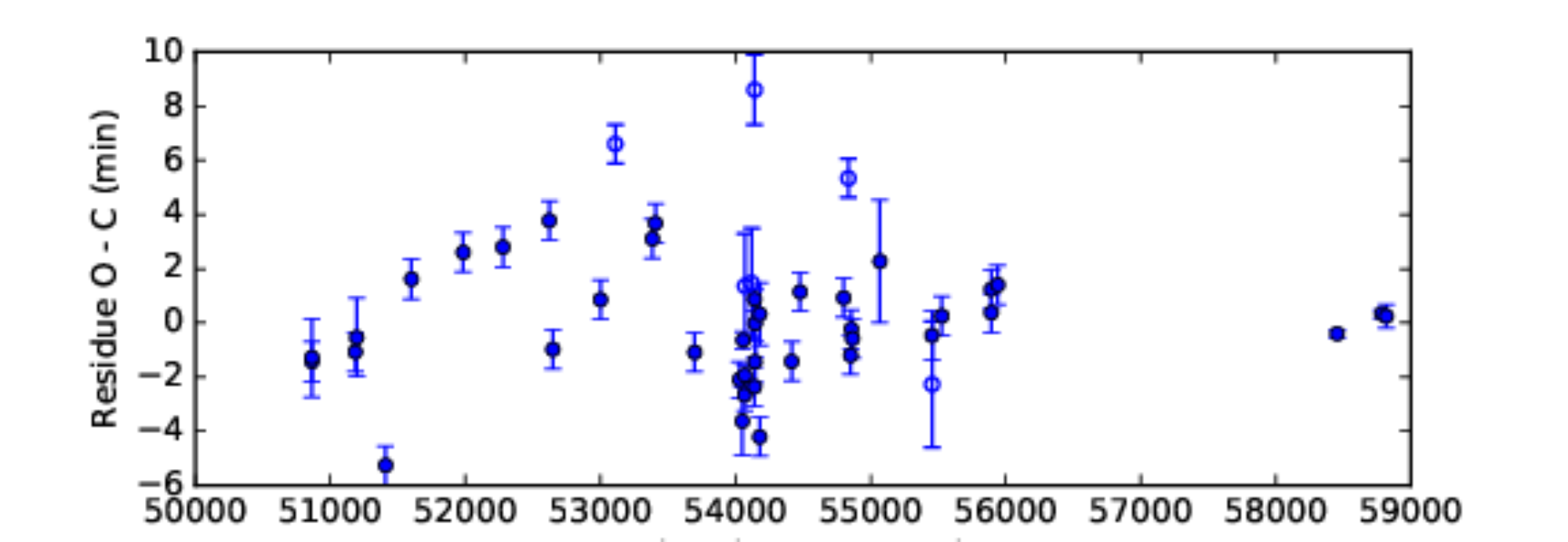}
	\includegraphics[trim=30 0 35 15, clip, width=0.46\textwidth]{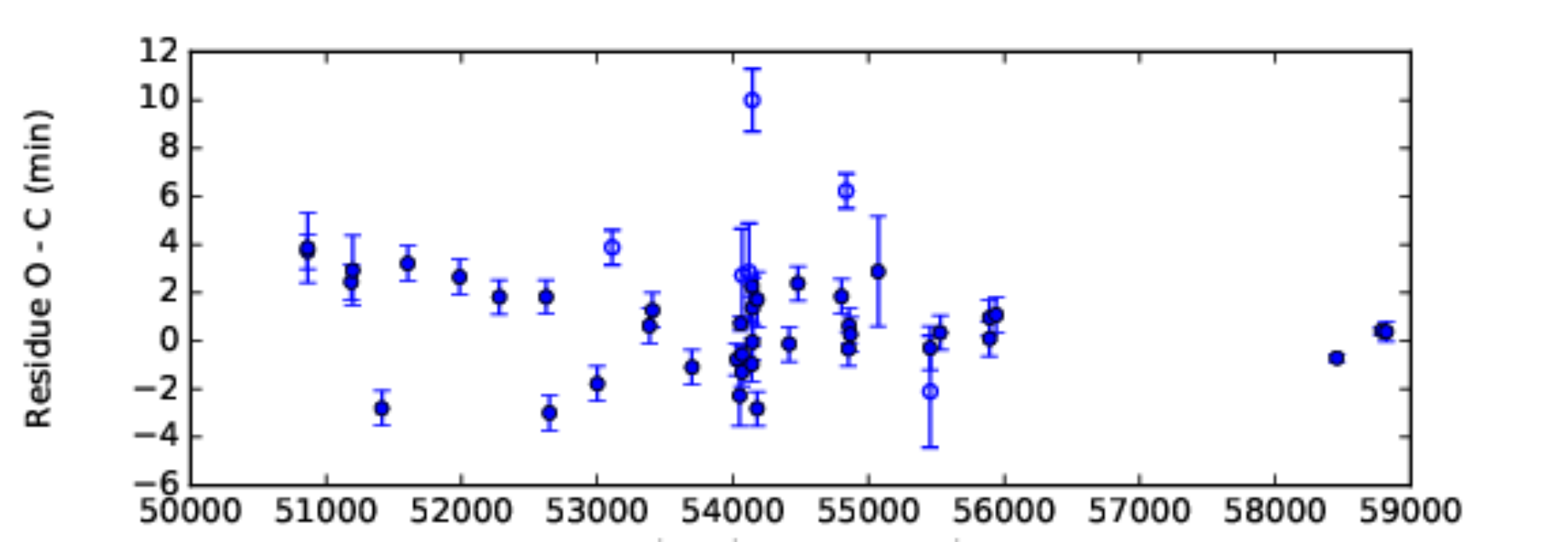}
	
	\caption{The fit (red line) of the O-C diagram of CR Tau systems according to a quadratic ephemeris (left) and 3$^{rd}$ body (right), with residuals at the bottom.}
	\label{fig:oc_CR_Tau}
\end{figure}

O-C diagram of CR Tau contains 48 CCD times of minima, including archival and new points. We again excluded old photographic minima times obtained by \cite{Hoffmeister1949}, because of their very large scatter (up to 2 hours on O-C diagram). The precision of CCD minima times is in the range of $10^{-4}$ days.  As in the previous case, we performed a weighted least-squares solution using all minima and determined the following linear ephemeris of the system:
\begin{equation}
  \mathrm {Min~I} = \mathrm {HJD~} 2452500.125(3) + 0^{d}.6827039(4) \times E, 
  \label{eq:lin_cr}
\end{equation}
which was used to create the O-C diagram depicted in Fig.~\ref{fig:oc_both}~(right). Unlike the previous case, there is some visible variation on O-C of CR Tau. Because of the lack of minima times and insufficient coverage we can only speculate about their nature. The first explanation can be a mass transfer between components in the system. We made a weighted least-squares solution of residuals and obtained quadratic ephemeris: 
\begin{equation}
  \mathrm {Min~I} = \mathrm {HJD~} 2452500.1267(12) + 0^{d}.6827026(6) \times E + 1.395(64) \times 10^{-10} \times E^2. 
  \label{eq:quad_cr}
\end{equation}
This solution is depicted in Fig.~\ref{fig:oc_CR_Tau}~(left). According to ephemeris (\ref{eq:quad_cr}), a long-term period increase at a rate of $1.494(8) \times 10^{-7} d/y$ is detected. The second possible explanation of the O-C diagram is the presence of the $3^{rd}$ body in the system, which we do not directly see. It causes a light-time effect, a shifting of minima times according to the movement of the visible binary around the common center of mass \citep{2001icbs.book.....H}. We have used code from \citet{2019OEJV..197...71G} to test this hypothesis. Our best solution (shown in  Fig.~\ref{fig:oc_CR_Tau}~right) led to the high eccentric $e=0.8$ orbit of the body with almost 28 years orbital period.  


\section{Light curve analysis}
\label{sec:lc_analysis}
For the analysis of light curves of both systems, we have used ELISa\footnote{\texttt{https://github.com/mikecokina/elisa}} code \citep{Cokina2020}. 
It is a newly developed cross-platform Python software package dedicated to modeling close eclipsing binaries including surface features such as spots and pulsations. ELISa utilizes modern approaches to the EB modeling with an emphasis on computational speed while maintaining a sufficient level of precision to process a ground-based and space-based observation. In this paper, we utilize its capability to model the light curves of close eclipsing binaries with the built-in capability to solve an inverse problem using the Least Squares (LS) and Markov Chain Monte-Carlo (MCMC) methods. 

\begin{figure}[!th]
	\centering
	\includegraphics[width=0.52\textwidth]{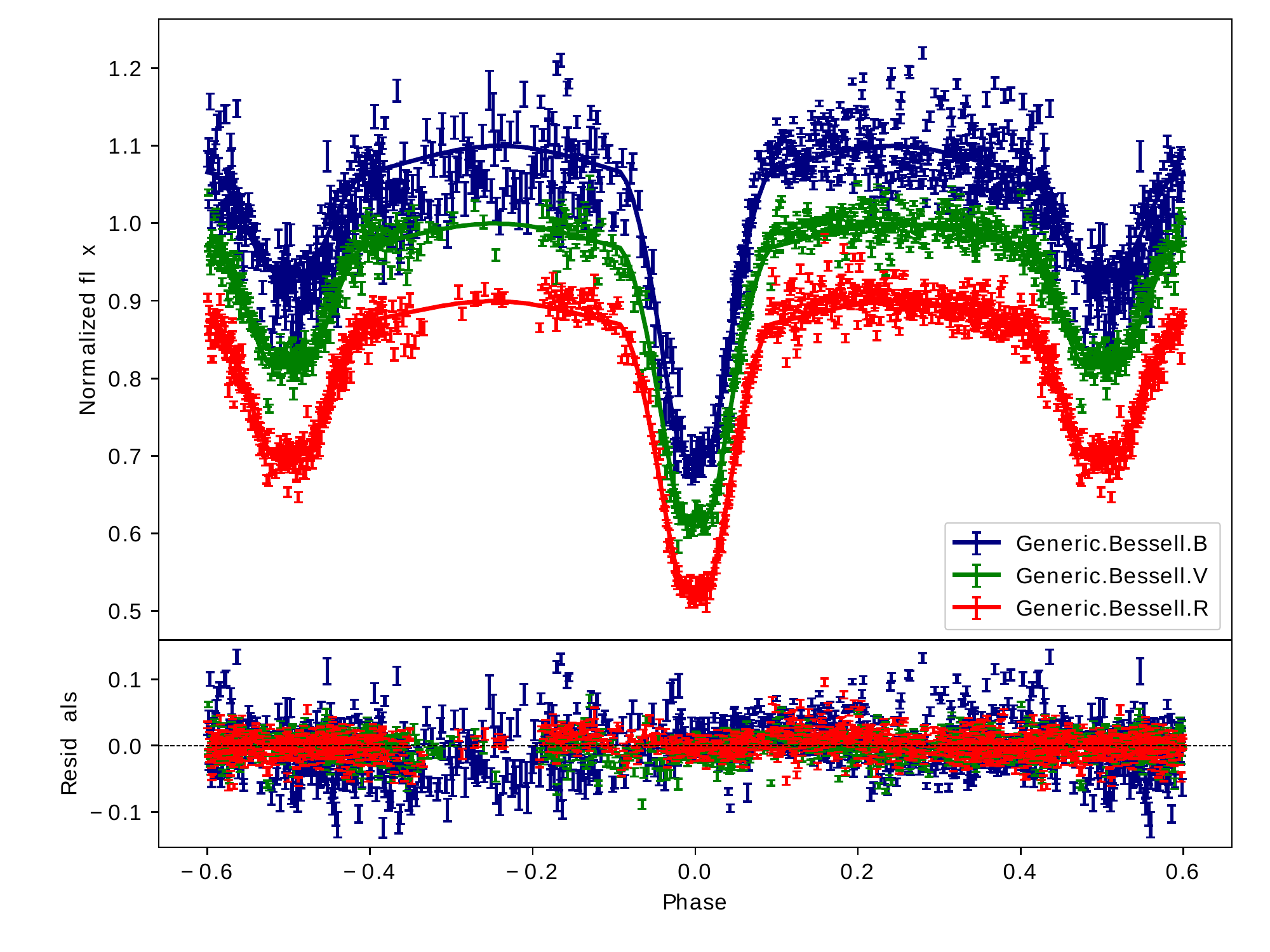}
	\includegraphics[width=0.45\textwidth]{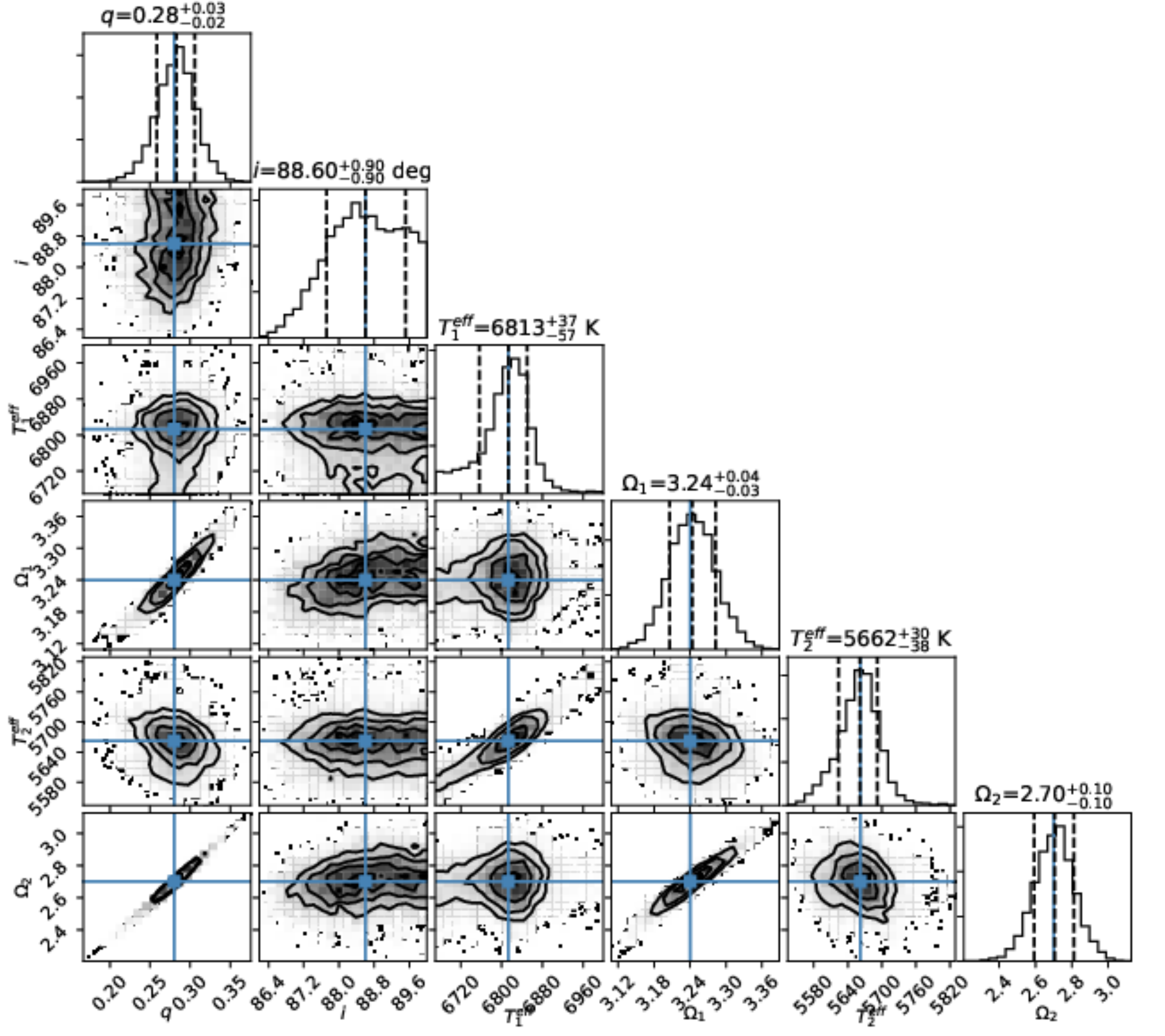}
	\caption{The synthetic model fitted on observational data of V1321~Cyg (left) and the results of the MCMC sampling displayed in form of the corner plot (right).}
	\label{fig:V1321_Cyg_all}
\end{figure}

\begin{figure}[th]
	\centering
	\includegraphics[width=0.52\textwidth]{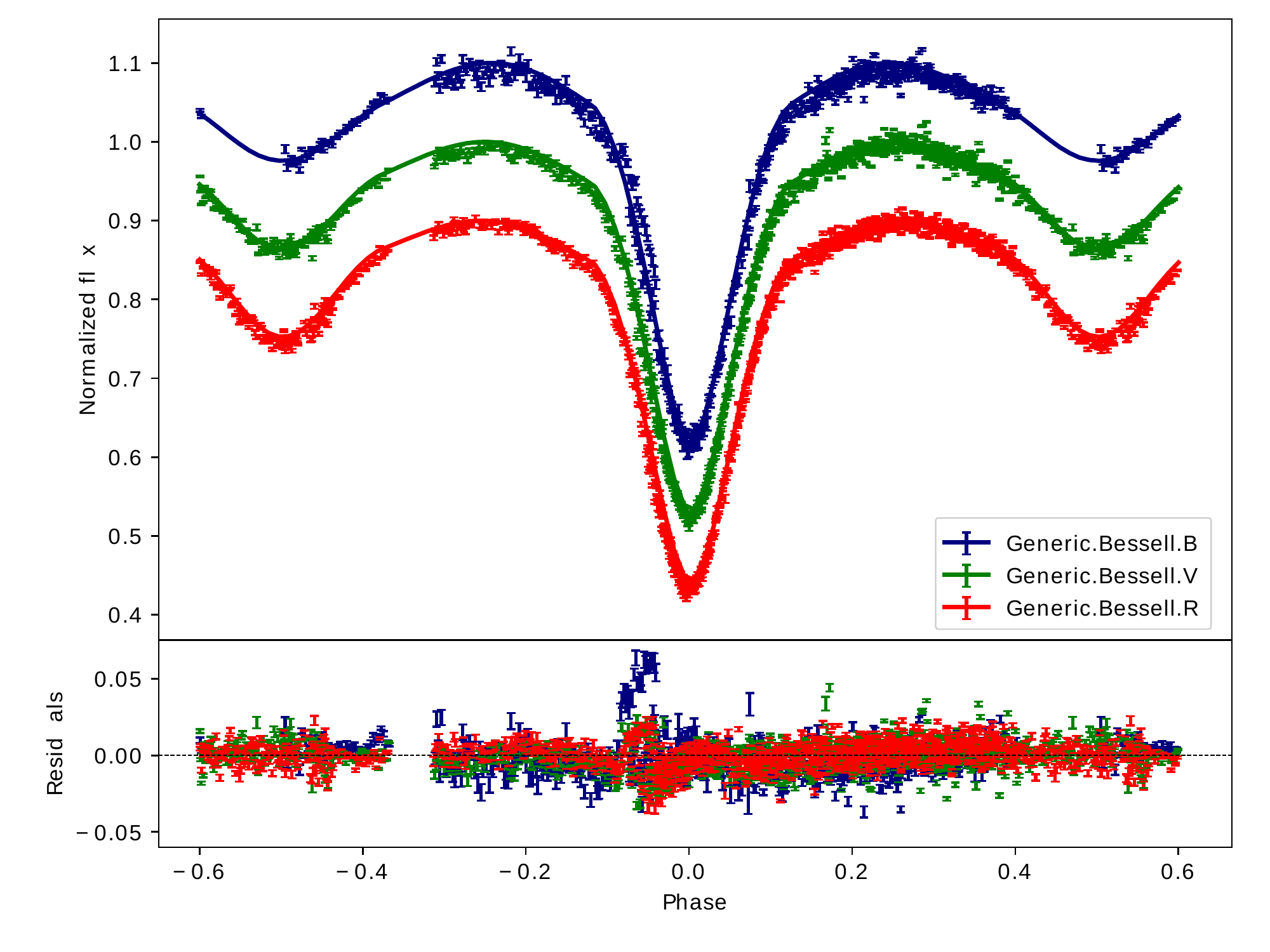}
	\includegraphics[width=0.45\textwidth]{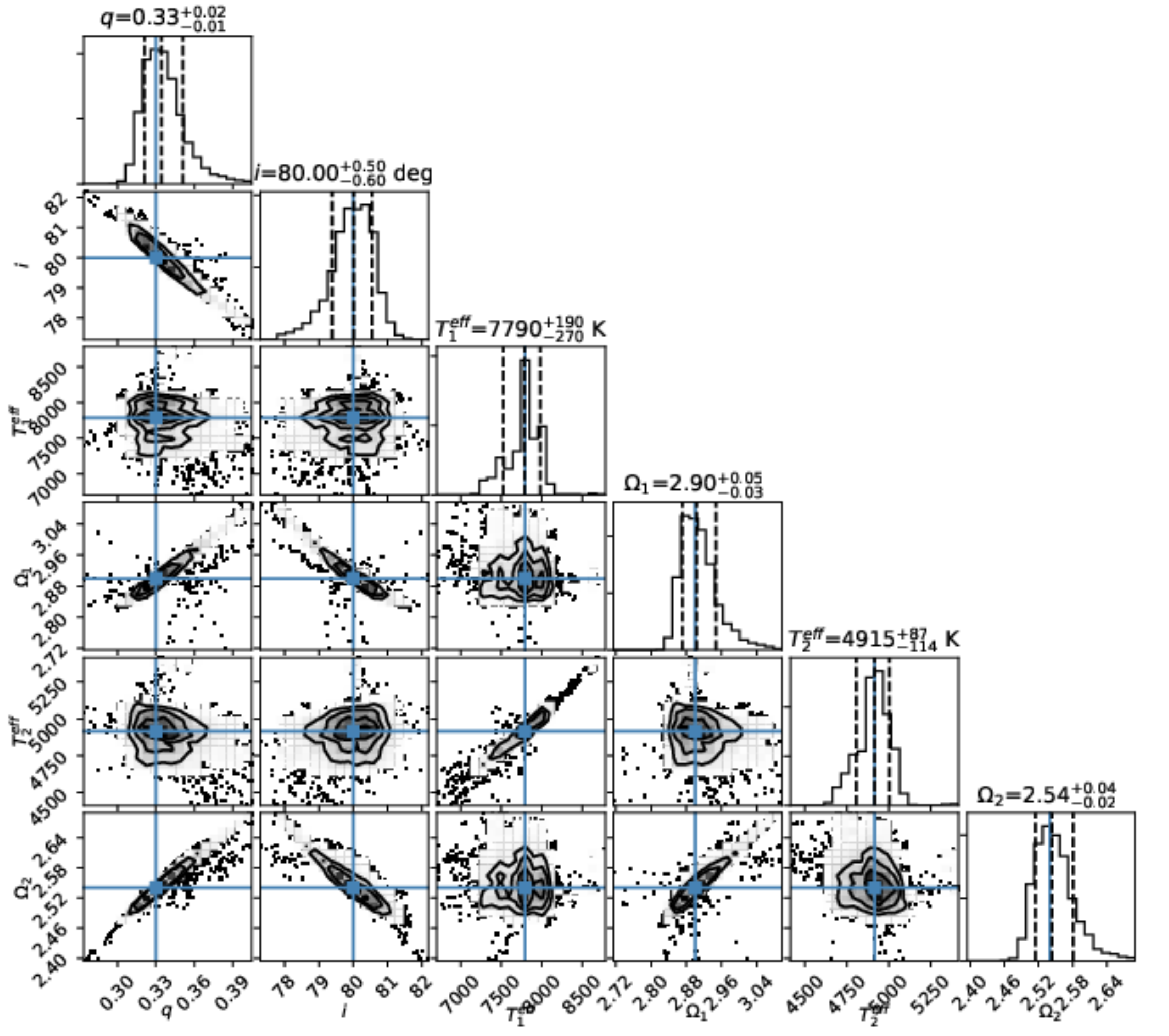}
	\caption{The synthetic model fitted on observational data of CR~Tau (left) and the results of the MCMC sampling displayed in form of the corner plot (right).}
	\label{fig:CR_Tau_all}
\end{figure}

\begin{figure}[!th]
	\centering
	\includegraphics[width=0.495\textwidth]{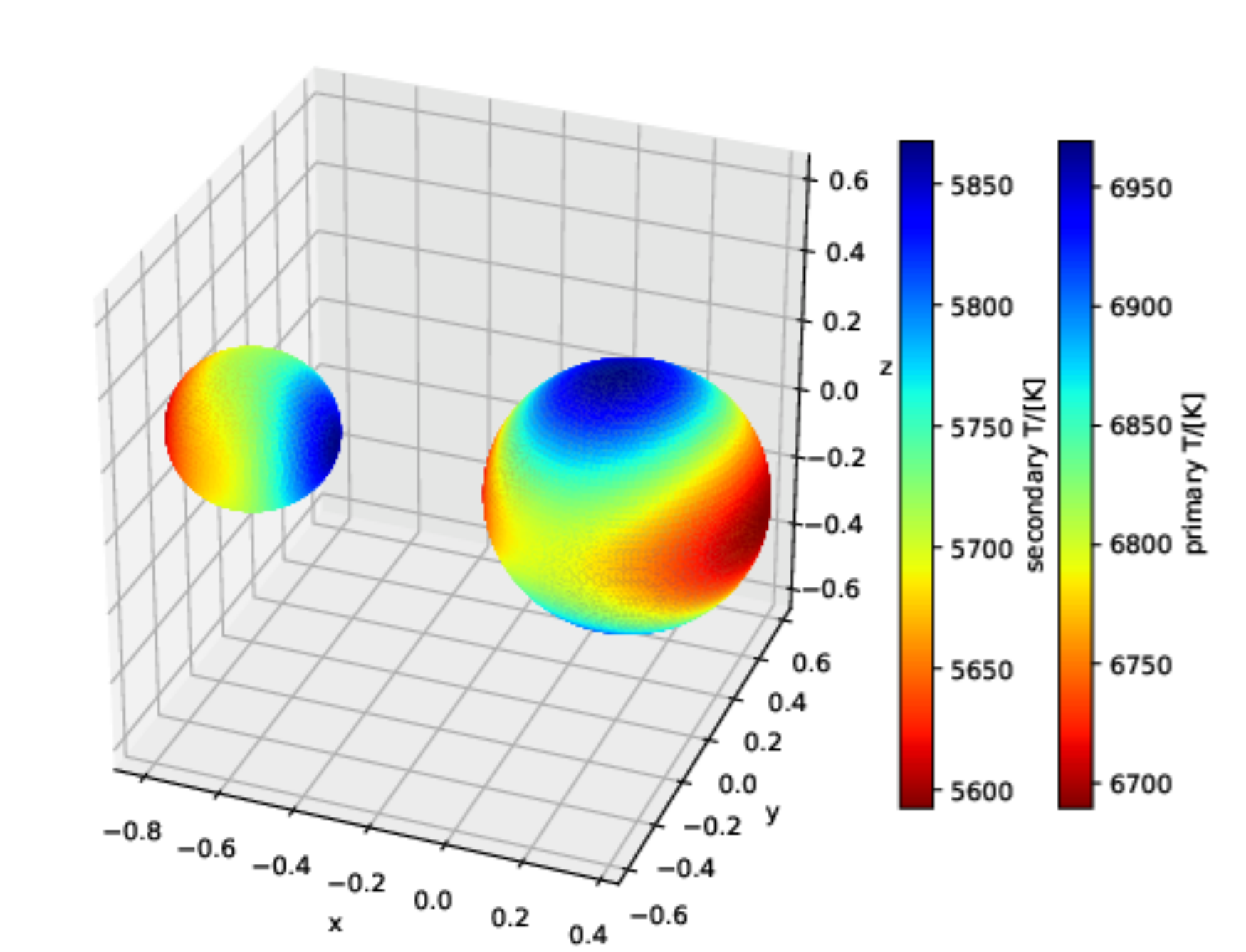}
	\includegraphics[width=0.495\textwidth]{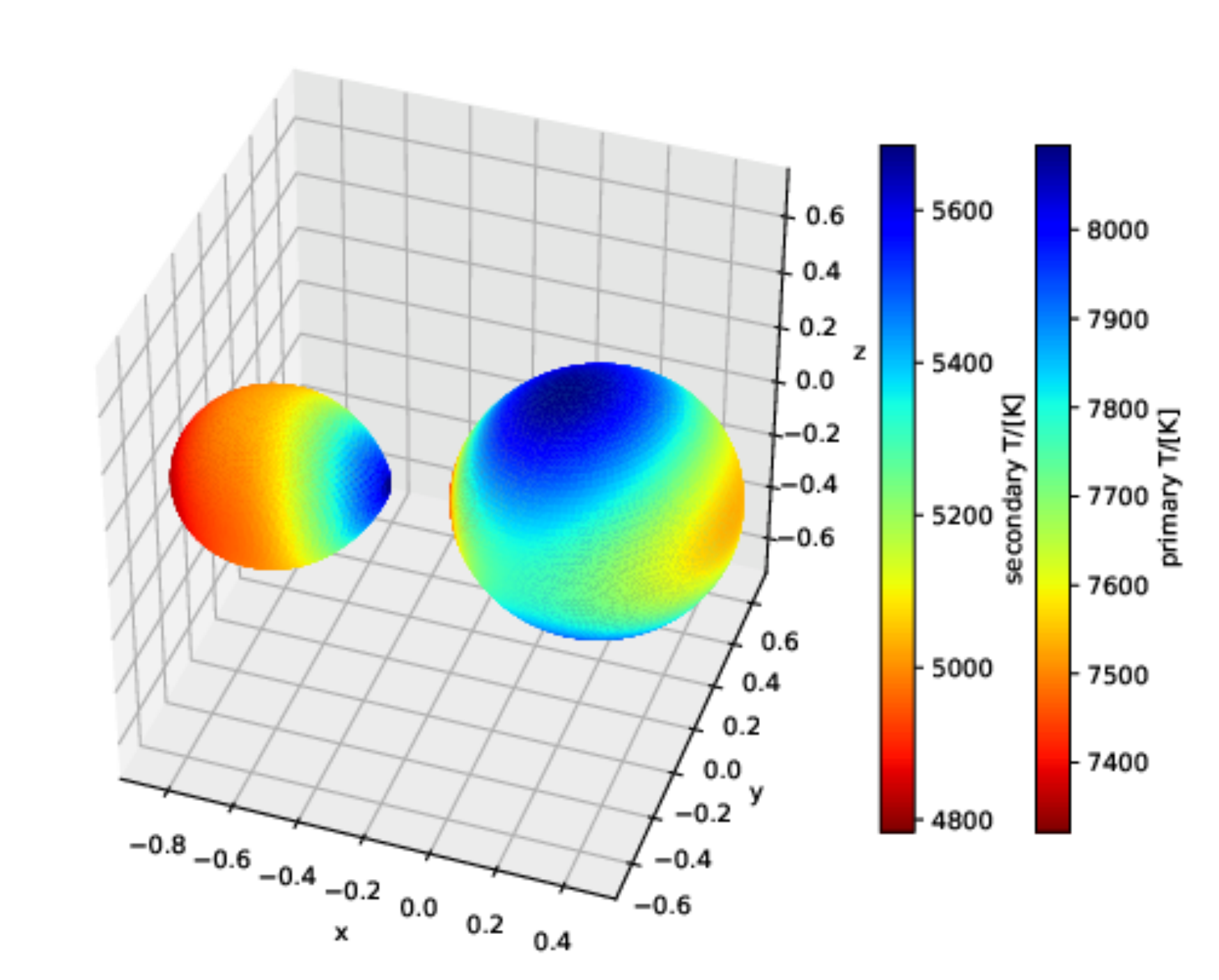}
	\caption{3D models with the most likely surface temperature distributions of V1321 Cyg (left) and CR Tau (right) binary systems based on parameters obtained from LC fitting listed in Tab.~\ref{tab:ph_param}. The displayed temperature distributions take into account gravity darkening and the reflection effect using standard techniques based on Wilson-Devinney code \citep{Wilson1971}.}
	\label{fig:CR_Tau_model}
\end{figure}

Observations in all passbands were normalized according to flux in the maxima and were simultaneously fitted by the LS method to find initial approximate solutions.
Subsequently, MCMC sampling was used to produce 1\,$\sigma$ confidence intervals of the fitted system parameters. Each system was fitted with model containing 6 free parameters: orbital inclination $i$, photometric mass ratio $q$, surface potentials of both components $\Omega_1$ and $\Omega_2$ and effective temperatures of the primary and secondary component $T^{eff}_1$ and $T^{eff}_2$. 
In case of V1321 Cyg,  $T^{eff}_1$ was kept during fitting procedure within $\pm 300$\,K of the value 6770\,K obtained from LAMOST spectra \citep{2018ApJS..235....5Q}. 
On the other hand, $T^{eff}_1$ of CR Tau was constrained within $\pm 1000$\,K interval from the value 8200\,K provided by the 2$^{nd}$ GAIA data release \citep{gaia2}.

For the components with convective envelopes (effective temperatures bellow $\sim$7000\,K), the albedos $A_1$, $A_2$ of components were set to 0.6 \citep{Rucinski1969}  and gravity darkening factors, $g_1$ and $g_2$ to 0.32 \citep{Lucy1967}. In the case of radiative envelope (above $\sim$7000\,K), the values of albedo and gravity darkening factor were both set to 1.0. \citet{2003IAUS..210P.A20C} models of stellar atmospheres were used. The linear limb darkening coefficients for each component were interpolated from the \citet{vanHamme1993} tables. 

The weights of individual data points were established as $1/\sigma^2$, where $\sigma$ is the standard error of point derived during photometric measurement. Initially, the LS algorithm was used with suitable initial parameters to find an approximate solution and then the parameter space near the solution was explored with MCMC sampler with 500 walkers and 500 iterations with prior 300 iterations discarded as it belonged to the thermalization stage of the sampling. 
The resulting and derived parameters of both systems, like a critical potential $\Omega_{crit}$, corresponding radius $R_{eqiv}$, and periastron radii in SMA are listed in Tab.~\ref{tab:ph_param}. The best-fit models with observed LCs and resulting flat chains displayed in the form of the corner plot are shown in Fig.~\ref{fig:V1321_Cyg_all} and \ref{fig:CR_Tau_all}. On the Fig.~\ref{fig:CR_Tau_model} we also display 3D models of the systems with a temperature distribution, corresponding to a best fitting solution listed in Tab.~\ref{tab:ph_param}.

\begin{table}[t]
\bc
\centering
\caption{Parameters of the V 1321~Cyg and CR~Tau systems derived from multi-color photometry using ELISa code. The description of parameters is given in text. Goodness of the obtained fits are provided in the form of coefficient of determination $R^2$. \label{tab:ph_param}} 
\setlength{\tabcolsep}{7pt}
\begin{tabular}{l||cc||cc}
\hline
\hline
\textbf{Parameter}    & \multicolumn{2}{c||}{\textbf{V1321 Cyg}}                            & \multicolumn{2}{c}{\textbf{CR~Tau}}                               \\
                      & Primary      &  Secondary              & Primary      &  Secondary   \\ 


                      
\hline                    
P [HJD]             & \multicolumn{2}{c||}{0.7281849$^a$}                                     & \multicolumn{2}{c}{0.6827039$^a$}                                     \\
T$_0$ [HJD]           & \multicolumn{2}{c||}{2458428.879$^a$}                                & \multicolumn{2}{c}{2452500.125$^a$}                               \\
i [deg]               &\multicolumn{2}{c||}{$88.6^{+0.9}_{-0.9}$}                           & \multicolumn{2}{c}{$80.0^{+0.5}_{-0.7}$}                       \\
q (M$_{2}$/M$_{1}$)   &\multicolumn{2}{c||}{$0.28^{+0.02}_{-0.03}$}                         & \multicolumn{2}{c}{$0.33^{+0.02}_{-0.01}$}                  \\
$T$ [K]               &   $6810^{+40}_{-60}$            & $5660^{+30}_{-40}$                &  $7790^{+190}_{-260}$            & $4916^{+87}_{-111}$              \\
$\Omega$              & $3.24^{+0.04}_{-0.04}$          & $ 2.7^{+0.1}_{-0.1}$              &  $2.90^{+0.04}_{-0.03}$          & $2.54^{+0.04}_{-0.03}$        \\
$\Omega_{crit}$       & $2.43^{+0.05}_{-0.06}$          & $ 2.43^{+0.06}_{-0.06}$           &  $2.54^{+0.04}_{-0.03}$          & $2.54^{+0.04}_{-0.03}$        \\
$R_{eqiv}$[SMA]       & $0.343^{+0.002}_{-0.002}$       & $ 0.214^{+0.002}_{-0.002}$        &  $0.401^{+0.003}_{-0.005}$       & $0.288^{+0.004}_{-0.003}$   \\ 
\hline                                                                                              
\multicolumn{5}{c}{\emph{Periastron radii [SMA]}}\\
\hline
$r_{polar}$           & $0.336^{+0.002}_{-0.002}$ &  $0.207^{+0.001}_{-0.001}$   &  $0.386^{+0.003}_{-0.004}$  & $0.2684^{+0.004}_{-0.003}$     \\
$r_{backward}$        & $0.350^{+0.002}_{-0.002}$ &  $0.222^{+0.003}_{-0.003}$   &  $0.414^{+0.003}_{-0.005}$  & $0.3117^{+0.004}_{-0.003}$     \\
$r_{side}$            & $0.345^{+0.002}_{-0.002}$ &  $0.211^{+0.002}_{-0.002}$   &  $0.402^{+0.004}_{-0.005}$  & $0.2794^{+0.004}_{-0.003}$     \\
$r_{forward}$         & $0.354^{+0.002}_{-0.002}$ &  $0.227^{+0.004}_{-0.004}$   &  $0.427^{+0.004}_{-0.005}$  & $0.376^{+0.008}_{-0.010}$      \\
\hline
$R^2$   &  \multicolumn{2}{c||}{0.932} & \multicolumn{2}{c}{0.994} \\         
\hline\hline
\end{tabular}
\ec
\tablecomments{0.74\textwidth}{$^a$ - adopted values of period and epoch from linear ephemerides}
\end{table}

\section{Discussion and conclusions}
\label{sec:discussion}
In our study we have presented the first multi-color $BVR$ photometry of two, so far neglected eclipsing binaries V1321~Cyg and CR~Tau. We have analyzed their period variations using archival and our new minima times as well as we performed a photometric analysis of their light-curves.

The analysis of our multi-color photometric observations demonstrates that V1321~Cyg is a close detached binary with a low photometric mass ratio of $q=0.28^{+0.02}_{-0.03}$. Such a low value of $q$  combined with a secondary potential $\Omega_2 = 2.7^{+0.1}_{-0.1}$ being relatively close to the critical potential $\Omega_{crit, 2} = 2.43^{+0.06}_{-0.06}$ suggests that V1321~Cyg is a post mass transfer system, where a significant portion of the secondary component's mass was transferred onto the primary component. We detected no significant period changes in this system and it also supports the idea that the system is detached. We found two viable solutions, however only one was located within the temperature range derived by the LAMOST spectra while the second discarded solution contained much colder primary component with $\approx 6000K$, well below the expected value.  

On the other hand, a photometric analysis of the light curves of CR Tau revealed that the system is a semi-detached system where the secondary component almost fills its Roche lobe, as detected in some other near-contact systems, like EG Cep 
\citep{Zhu2009} 
or BF Vir 
\citep{Zhu2012} 
The main consequence of such configuration is a mass transfer from the secondary to the primary component, which is reflected on the O-C diagram as a parabolic variation according to the epoch. If the mass is transferred from a less massive star to a more massive one, we detect period increase, as observed in our data (see Fig.~\ref{fig:oc_CR_Tau} - left). So we can conclude that the most probable explanation of O-C variations of CR~Tau is a mass transfer and further observations should confirm that.

Subtraction of the best fit from the observed multi-colour data (Fig.~\ref{fig:CR_Tau_all} - left) uncovered phase correlated residuals centered around the primary eclipse. This can be also explained by the mass transfer from the less massive and much cooler secondary component onto the heavier primary component. 
The surface of the primary component is obscured by the fraction of the relatively cold stream of matter from the secondary component. A slight shift in the position of residuals to the beginning of the eclipse can be explained by the Coriolis force acting on the falling stream of matter. Additionally, it is worth to mention other observed proximity effects such as the deformation and heating of secondary component of CR Tau on the part of the surface facing the primary component due to the close proximity of the components and large temperature difference between the component's surfaces.

\normalem
\begin{acknowledgements}
This work was supported by national grant 0119U100236 and by the Slovak Research and Development Agency under contract No. APVV-15-0458. The research of M.F. was supported by the internal grant No. VVGS-PF-2019-1392 of the Faculty of Science, P. J. {\v S}af{\'a}rik University in Ko{\v s}ice. 

\end{acknowledgements}

\bibliographystyle{raa}
\bibliography{ms0398}

\end{document}